\documentclass[]{spie}  

\usepackage{amsmath,amsfonts,amssymb}
\usepackage{graphicx}
\usepackage[colorlinks=true, allcolors=blue]{hyperref}
\usepackage{comment}
\usepackage{subcaption}
\usepackage{wrapfig}

\usepackage{url}
\usepackage{hanging}
\usepackage{lipsum}
\usepackage{multicol}
\usepackage{multirow}
\usepackage{overpic}
\usepackage[inline]{enumitem}
\usepackage{setspace}
\usepackage[labelfont=bf, singlelinecheck=false, justification=justified]{caption}
\usepackage{wrapfig}
\usepackage{comment}
\usepackage{multicol}
\usepackage[normalem]{ulem}
\usepackage{parskip}  
\usepackage{multirow}
\usepackage[table,xcdraw,dvipsnames]{xcolor}
\usepackage[linewidth=0.5pt,skipabove=10pt,skipbelow=10pt,leftmargin=0pt,rightmargin=0pt,framemethod=TikZ]{mdframed}
\definecolor{myblue}{RGB}{25, 118, 210}
\mdfdefinestyle{bluebox}{backgroundcolor=myblue!10,roundcorner=8pt,linecolor=myblue}
\usepackage{pdfpages}  
\usepackage{soul,color} 

\title{The Focal-plane Actualized Shifted Technique Realized for a Shack Hartmann Wavefront Sensor (fastrSHWFS)}

\author[a]{Benjamin L. Gerard}
\author[a]{Aaron Lemmer}
\author[a]{Bautista R. Fernandez}
\author[a]{Xiaoxing Xia}
\author[a]{Cesar Laguna}
\author[a]{Mike Kim}
\author[a]{Stephen Mark Ammons}
\author[a]{Brian Bauman}
\author[a]{Lisa Poyneer}
\affil[a]{Lawrence Livermore National Laboratory}

\authorinfo{Further author information. Send correspondence to Benjamin L. Gerard: \texttt{gerard3@llnl.gov}
}
\begin{document} 
\maketitle

\begin{abstract}
Astronomical adaptive optics (AO) is a critical approach to enable ground-based diffraction-limited imaging and high contrast science, with the potential to enable habitable exoplanet imaging on future extremely large telescopes. However, AO systems must improve significantly to enable habitable exoplanet imaging. Time lag between the end of an exposure and end of deformable mirror commands being applied in an AO loop is now the dominant error term in many extreme AO systems (e.g., Poyneer et al. 2016), and within that lag component detector read time is becoming non-negligible (e.g., Cetre et al. 2018). This term will decrease as faster detector readout capabilities are developed by vendors. In complement, we have developed a modified Shack Hartmann Wavefront Sensor (SHWFS) to address this problem called the Focal-plane Actualized Shifted Technique Realized for a SHWFS (fastrSHWFS). The novelty of this design is to replace the usual lenslet array with a bespoke pupil-plane phase mask that redistributes the spot pattern on the detector into a rectangular array with a custom aspect ratio (in an extreme case, if the detector size can accommodate it, the array can be a single line). We present the fastrSHWFS concept and preliminary laboratory tests. For some detectors and AO systems, the fastrSHWFS technique can decrease the read time per frame compared to a regular SHWFS by up to 30x, supporting the goal of reduced AO lag needed to eventually enable habitable exoplanet imaging. 
\end{abstract}


\section{Introduction}
\label{sec:intro} 
The recent 2020 Decadal Survey of Astronomy and Astrophysics\cite{NASEM2021} listed habitable exoplanet imaging with future extreme adaptive optics (AO) on 30m-class telescopes as a key priority in the coming decade. However, there is a current 100x contrast gap between the current state-of-the-art and what is needed to enable this goal\cite{Jensen-Clem2022}. Temporal error is generally the dominant error budget term for extreme AO systems (e.g., Ref.~\citenum{Poyneer2016}) contributing to this 100$\times$ gap. There are several strategies to reduce this error: (1) run the AO system faster (reducing the half-frame delay), (2) reduce the computational latency with faster computers and/or algorithms, (3) implement predictive control algorithms, and/or (4) reduce the hardware latency of the wavefront sensor camera and/or deformable mirror. Here we take the fourth approach with a solution we call the The Focal-plane Actualized Shifted Technique Realized for a Shack Hartmann Wavefront Sensor (fastrSHWFS), which changes the aspect ratio of the spot pattern so that it occupies fewer rows of the detector, reducing the read time and, in turn, the total system latency.
\section{Concept}
\label{sec:concept}
\begin{wrapfigure}[41]{r}{0.66\textwidth}
    \centering
    \begin{subfigure}[t]{0.65\textwidth}
        \includegraphics[width=1.0\textwidth]{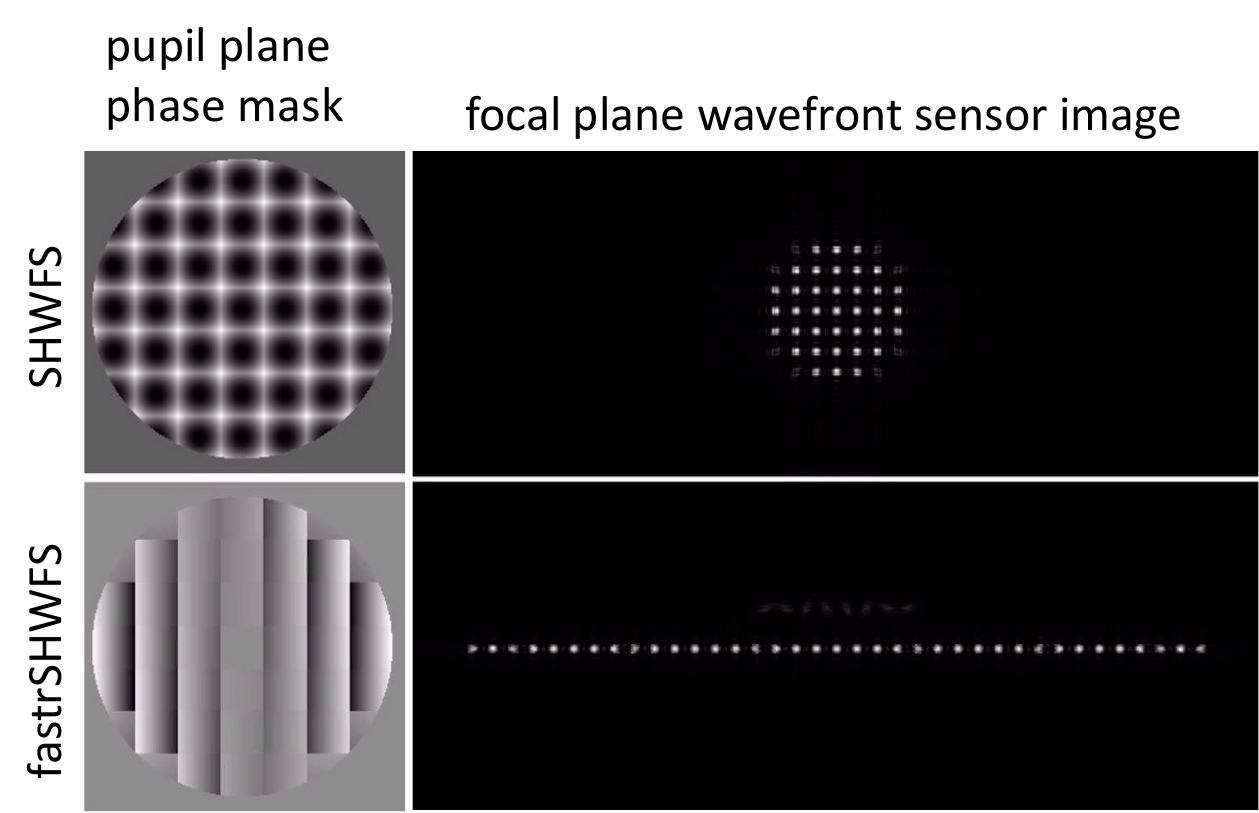}
        \caption{SHWFS (top row) vs.\ fastrSHWFS (bottom row) concept for a setup with 7 subapertures across an unobscured circular pupil, showing the corresponding pupil plane mask (left column) and resultant focal plane image (right column).}
        \label{fig:concept-a}
    \end{subfigure}
    \begin{subfigure}[t]{0.65\textwidth}
        \includegraphics[width=1.0\textwidth]{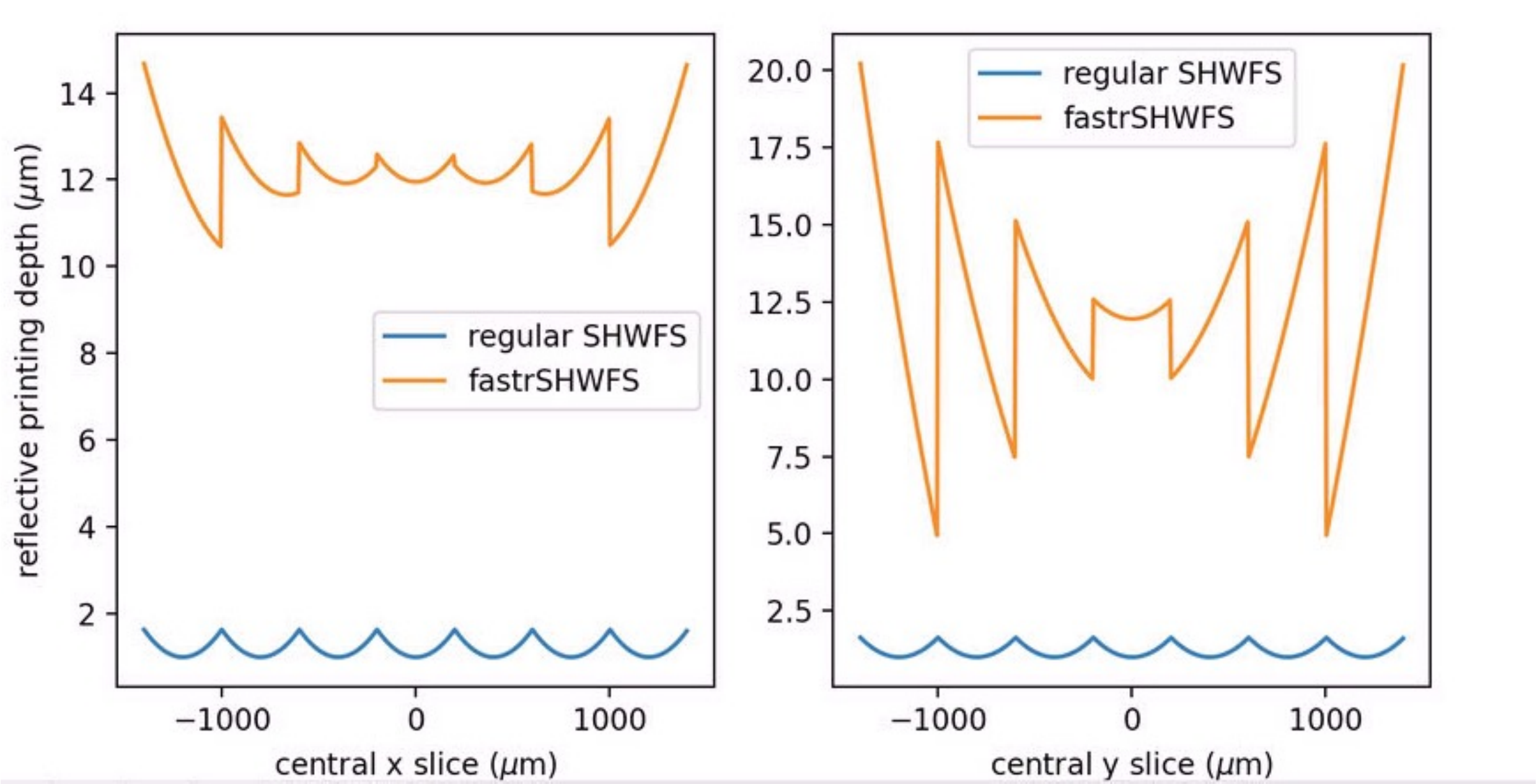}
        \caption{For the same configuration as panel (a), this shows the optical path difference in scalar printing or etching depth for a reflective mask for both a conventional lenslet array (with a 16 mm lenslet focal length and 400 $\mu$m lenslet pitch) with a fastrSHWFS mask (producing subapertures along a line separated by 2~$\lambda/d$).}
        \label{fig:concept-b}
    \end{subfigure}
    \vspace*{3pt}
    \caption{fastrSHWFS concept illustration.}
    \label{fig:concept}
\end{wrapfigure}

Fig.~\ref{fig:concept} illustrates the fastrSHWFS concept. In short, the technique leverages a custom lenslet array mask to redistribute subapertures from a circular pupil geometry into a rectangular or linear geometry on the wavefront sensor (WFS) detector, leveraging the principle that detectors generally have a shorter read time when reading out less rows. Fig.~\ref{fig:concept-a} shows that spots below an illumination threshold (usually 50\% relative to a fully illuminated subaperture) can be optically redistributed to another line (in this figure, above the main row of fully illuminated subapertures) to save space on the region of the detector intended for wavefront sensing. A secondary obscuration and support structures can also be accounted for in the design to similarly only place greater than 50\% illuminated subapertures in the linear/rectangular region of interest for real-time wavefront sensing and control.

A fastrSHWFS mask can be fabricated two ways: (1) with a focus and tip/tilt component included in each subaperture or (2) with just a tip/tilt component included in each subaperture. In principle, option~1 needs no reimaging optics between the fastrSHWFS mask and the WFS detector (although in practice optomechanically this may be challenging and require one or more re-imaging lenses). Option~2 does inherently require a powered optic downstream from the fastrSHWFS mask, where the mask forms ``beamlets'' on that powered optic (and potentially also reimaging lenses for the same reason as Option~1) which then images subapertures onto the WFS detector. Fig.~\ref{fig:concept-b} shows the required etching/printing depths for Option~1 for a fastrSHWFS with seven subapertures across the beam diameter, needing greater than 20~$\mu$m depths for reflective masks. For a system with thirty subapertures across the beam diameter, greater than 200~$\mu$m of etching or printing depth is needed for a reflective mask and, correspondingly, $2/(n-1)$ times greater depth for a transmissive mask (where $n$ is the material's index of refraction). Tradeoffs between reflective and transmissive masks and the designs with or without intrinsic focus are apparent, all of which we explore later in this paper.
\newline
\section{Motivation}
\label{sec:motivation}
\begin{wrapfigure}[18]{r}{0.4\textwidth}
    \centering
    \captionof{table}{fastrSHWFS read time implications for three different COTS cameras. ROI read times and scaling laws with the number of rows and columns were obtained from each respective camera vendor (private communication). Listed read times illustrate the fastrSHWFS concept's potential benefits.}
    \label{tab:cots_cameras}
    \includegraphics[width=0.4\textwidth]{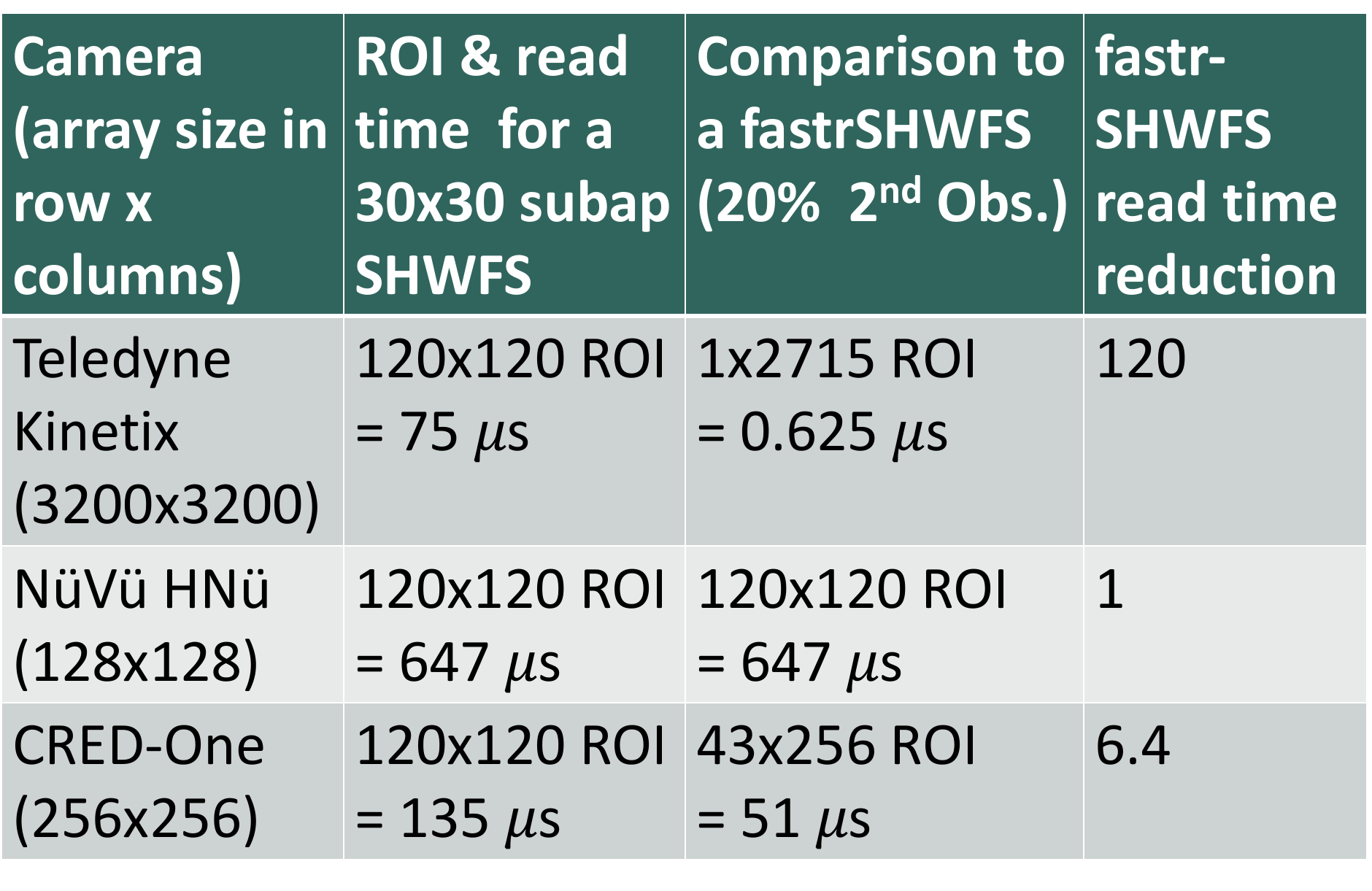}
\end{wrapfigure}

Table \ref{tab:cots_cameras} shows examples of commercial-off-the-shelf (COTS) cameras and how the fastrSHWFS would contribute (or not) to reduced latency from the use of these cameras by reading out different regions of interest (ROIs). Using the C-RED One results of Table~\ref{tab:cots_cameras}, we illustrate how this reduced read time translates into AO bandwidth increase and temporal error reduction. Using Ref.~\citenum{Cetre2018}, a Pyramid wavefront sensor system using a C-RED One (but for the purposes of this argument, assume the system is a SHWFS), the total system latency when running at an AO loop rate of 1~kHz is 1.636~ms, which includes the WFS exposure half frame delay; within that latency, the read time accounts for 709~$\mu$s. The system in Ref.~\citenum{Cetre2018} has developed custom C-RED One readout electronics and software that differs from the baseline read times for the commercially available C-RED One cameras, but because this system uses a 128$\times$128 ROI and the detector is 
256$\times$320 (rows $\times$ columns), from Table \ref{tab:cots_cameras} we assume the potential for reduced read time is also a factor of 6.4 (which is calculated assuming there are 256 columns, so the actual read time reduction should be more than this), assuming despite the different absolute read times that they have the same read time scaling law with pixel rows as the COTS C-RED One. Thus, a fastrSHWFS system read time would be reduced 6.4$\times$ from 709~$\mu$s to 111~$\mu$s, providing a total system latency of 1.0377 ms (again, including the half-frame delay). Using the AO transfer functions from Ref.~\citenum{Alloin1994} and to ensure loop stability enforcing phase and gain margins of 45$^\circ$ and 2.5, respectively, this read time reduction increases the 0-dB bandwidth from 66~Hz (gain = 0.5, leak = 0.99) to 100~Hz (gain = 0.78, leak = 0.99), a 52\% bandwidth increase. From Ref.~\citenum{Hardy1998} (Eqn.~9.53), this translates into a decrease of the temporal error term in the AO error budget by a factor of 1.4. Per Ref.~\citenum{Poyneer2016}, at $\sim$6 $\lambda/D$ for the Gemini Planet Imager extreme AO system this 1.4$\times$ bandwidth error reduction corresponds to a $\sim$2$\times$ raw contrast reduction. Per Ref.~\citenum{Bailey2016} this should also decrease final contrast by 2$\times$ at the same $\sim$6 $\lambda/D$ separation.

\section{Preliminary Laboratory Results}
\label{sec:testing}
We have designed, fabricated, and tested both transmissive and reflective fastrSHWFS mask designs with and without focus, discussed further in the subsections below. Testing was done on sub-benches of the Low-Latency Adaptive Optical Mirror System (LLAMAS) testbed at Lawrence Livermore National Laboratory (LLNL).\cite{Ammons2018, Poyneer2023} The testing presented in this paper was done to characterize fastrSHWFS mask quality and was not implemented in closed-loop with the rest of LLAMAS AO hardware. All image characterization here was done using a CMOS camera with 5.86-$\mu$m pixel pitch.
\subsection{Zeiss Reactive Ion Etching-based Masks}
\label{sec:Zeiss}
\begin{figure}[!t]
    \centering
    \begin{subfigure}[t]{0.47\textwidth}
        \centering
        \includegraphics[height = 2.1in]{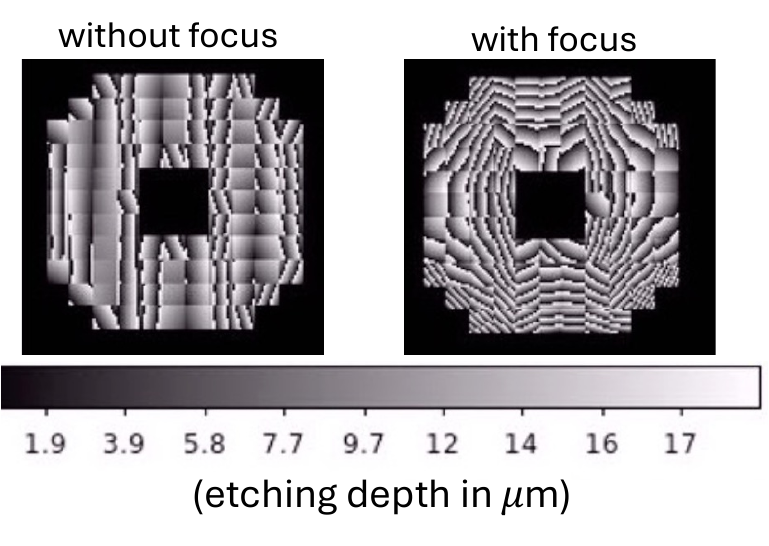}
        \caption{Zeiss' phase-wrapped ($\lambda$ = 355~nm) designs (11~subaps across each pupil converted into a 3x28 subap rectangle), showing etching depth into fused silica (limited to $<20\mu$m).}
        \label{fig:zeiss-a}
    \end{subfigure}
    \hspace{0.02\textwidth}
    \begin{subfigure}[t]{0.47\textwidth}
        \centering
        \includegraphics[height = 2.1in]{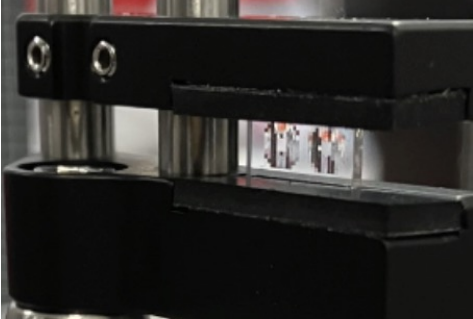}
        \caption{Fabricated masks without focus mounted in our optical system (substrate length x width x height = 5 x 10 x 1 mm). One pupil is a backup mask. Masks with focus have equivalent substrate dimensions and mounting setup.}
        \label{fig:zeiss-b} 
        \vspace{6pt}
    \end{subfigure}
    \begin{subfigure}[t]{1.0\textwidth}
        \includegraphics[width=1.0\textwidth]{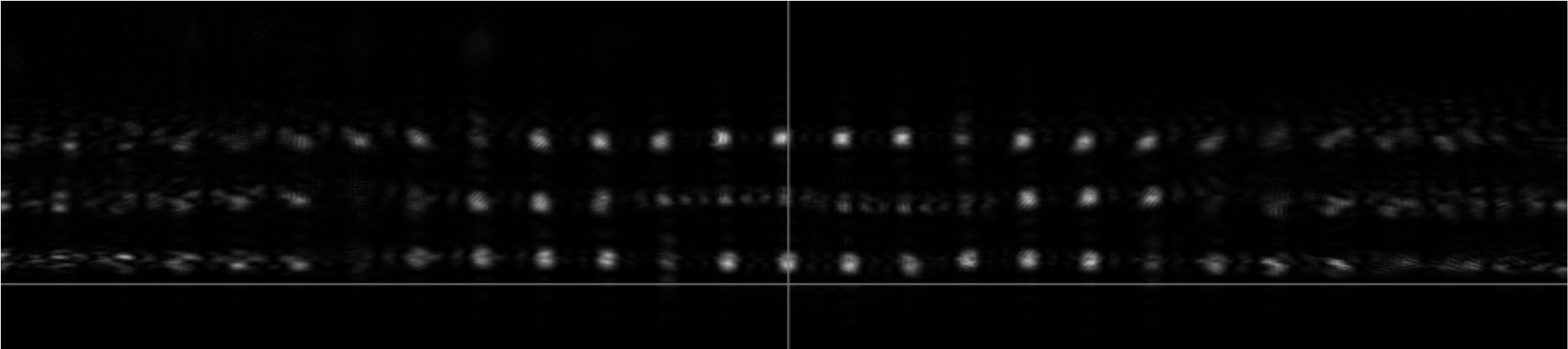}
        \caption{\centering Testing results using a Zeiss fastrSHWFS mask without focus.}
        \label{fig:zeiss-c}
        \vspace{6pt}
    \end{subfigure}
    \begin{subfigure}[t]{1.0\textwidth}
        \includegraphics[width=1.0\textwidth]{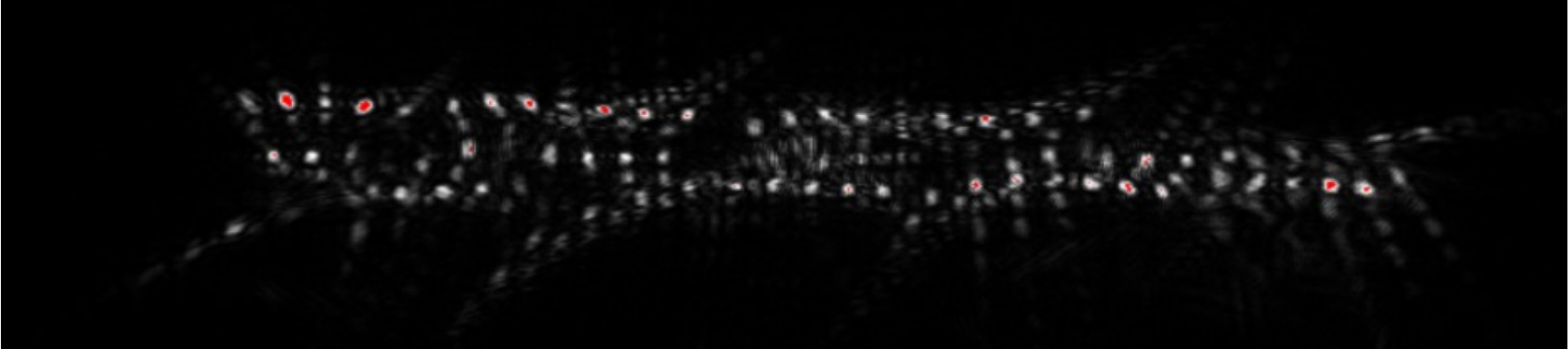}
        \caption{\centering Testing results using a Zeiss fastrSHWFS mask with focus.}
        \label{fig:zeiss-d}
        \vspace{6pt}
    \end{subfigure}
    \caption{Zeiss fastrSHWFS design and testing results.}
    \label{fig:zeiss}
\end{figure}

We designed and produced custom fastrSHWFS masks from Zeiss, fabricated by a lithographic etching process described in Ref.~\citenum{Cumme2015}. The design, fabricated masks, and testing results are shown in Fig.~\ref{fig:zeiss}.
As shown in Fig.~\ref{fig:zeiss-a}, these are transmissive masks designed for etching into fused silica, where the design is monochromatic for $\lambda=355$~nm due to the $<20$ $\mu$m etching depth limitation of Zeiss' lithographic fabrication technique. Phase-unwrapped designs with and without focus would require 120~$\mu$m and 50~$\mu$m maximum etching depths, respectively. These transmissive masks are beneficial in enabling simple comparison with COTS lenslet arrays already on LLAMAS, and they do not suffer from potential beam ellipticity and/or distortion errors inherent to a reflective mask, discussed next in \S\ref{sec:nanoscribe}. However, these phase-wrapped designs have disadvantages of (1) being inherently monochromatic and thus not as applicable to on-sky natural guide star AO observations and (2) Zeiss' etching technique producing depth-dependent errors, resulting in higher fabrication errors for the higher angle subapertures (i.e., those further from the optical axis). We anticipated these errors for the masks without focus, simulating that $\pm$5\% depth errors, a requirement which Zeiss said they could meet, would provide sufficient optical quality to produce diffraction-limited spots in our intended 3$\times$28 subaperture geometry. We were not, however, able to simulate the effect of $\pm$5\% depth errors on the mask with focus, as not even PROPER\cite{Krist2007} can simulate a coherent SHWFS, instead requiring incoherent simulation one subaperture at a time (J. Krist, private communication), so instead we decided to accept this tolerance requirement and see what testing results revealed.

As shown in Fig.~\ref{fig:zeiss-c}-\ref{fig:zeiss-d}, testing results measure insufficient quality to produce diffraction-limited spots for either mask type. Our testing setup included an adjustable pupil mask whose secondary obscuration fraction matched to the fastrSHWFS design and was conjugated to the fastrSHWFS mask via visible alignment with a shear plate and by minimizing Fresnel-ringing via camera imaging at 355~nm. The pupil mask had XY micrometer adjustability to allow co-alignment with the fastrSHWFS at the $\lesssim$10\% of a lenslet pitch level. Reimaging optics were carefully designed and tested to not vignette the post-lenslet fastrSHWFS beam, which is highly divergent given the large angles inherent in the design (largest ``beamlet'' angles of order 5$^\circ$). The camera was also mounted on a focus micrometer stage aligned to the optical axis to accurately find the best focus for all the spots (which have a designed depth of focus of order $\sim$1~mm). The mask without focus (Fig.~\ref{fig:zeiss-c}) did properly display all spots in the correct orientation, but the higher angle spots that were more heavily phase-wrapped were not of sufficient optical quality, which we confirmed by translating the re-imaging lens in X and seeing corresponding X motion of the spots but not an improvement of off-axis or on-axis spots relative to the image shown in the figure (if there was, that would suggest distortion from the re-imaging lens, but instead these tests clearly demonstrated that the fastrSHWFS mask itself was of insufficient quality for the highest angle subapertures). The dark area in the center of Fig.~\ref{fig:zeiss-c} is from the spiders in the pupil mask, which had to be fabricated at 33\% of a subaperture to spatial scale limitations of the laser etching technique used to fabricate the pupil mask (i.e., requiring $>100\; \mu$m-wide spiders). We were not able to obtain good quality spots for the fastrSHWFmask with focus (Fig.~\ref{fig:zeiss-d}), likely due to the combination of the increased phase wrapping (and associated above-discussed depth-dependent errors) needed for this design and the above-discussed inability to simulate mask design tolerances (e.g., it is possible that masks needed to reach diffraction-limited spot quality are an unrealistic fabrication error tolerance that Zeiss cannot reach). If fabrication errors tolerances are not feasable for future transmissive masks, reflective masks with this etching technique could be considered, which would relax the etching depth requirements by 5$\times$, although likely ultimately still requiring phase-wrapping for the highest aspect ratio fastrSHWFS geometries due to the $<20$~$\mu$m etching depth limitation of this micro-fabrication technique.
\subsection{Nanoscribe 3D Printed Masks}
\label{sec:nanoscribe}
We also tested fastrSHWFS mask fabricated by Nanoscribe 3D printing, a commercially available micro-3D printing machine via two-photon polymerization\cite{Baldacchini2016} that we have in-house at LLNL. The potential advantages of this microfabrication approach is that the technique has the ability to print hundreds of microns of printing depths that allows for unwrapped, achromatic designs and the application of the more linear geometries over a higher number of subapertures across the beam. However, this technique requires printing a resin material that is not optimal for optical transmission, thus requiring the masks to be coated and reflective. Reflective masks can constrain the opto-mechanical layout, and require a reflection angle that creates an elliptical beam on the fastrSHWFS lenslet mask as the reflection angle increases, which adds error to both the intended lenslet illumination pattern and for the mask with focus adds optical errors to the designed on-axis parabolic focus component reflected off-axis. These reflection angle effects could in principle be compensated for in design (i.e., create rectangular lenslets and design the focus component as off-axis parabolas instead of on-axis parabolas), but we did not do this for our masks tested in this work. Another drawback of the Nanoscribe 3D printing approach is that it requires stitching during the printing process because the continuous printing area is limited to less than 400 $\mu$m $\times$ 400 $\mu$m, although we have addressed this in our design by making the subaperture edges align with this limit.

\begin{figure}[b]
    \centering
    \begin{subfigure}[t]{0.17\textwidth}
        \includegraphics[width=1.0\textwidth]{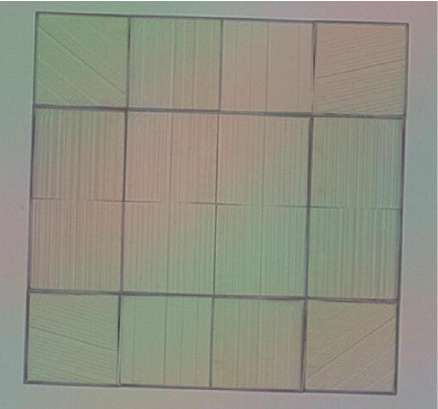}
        \caption{Microscope image of mask without focus (100~$\mu$m lenslet pitch).}
        \vspace{3pt}
        \label{fig:nanoscribe-a}
    \end{subfigure}
    \hspace*{2pt}
    \begin{subfigure}[t]{0.8\textwidth}
        \includegraphics[width=1.0\textwidth]{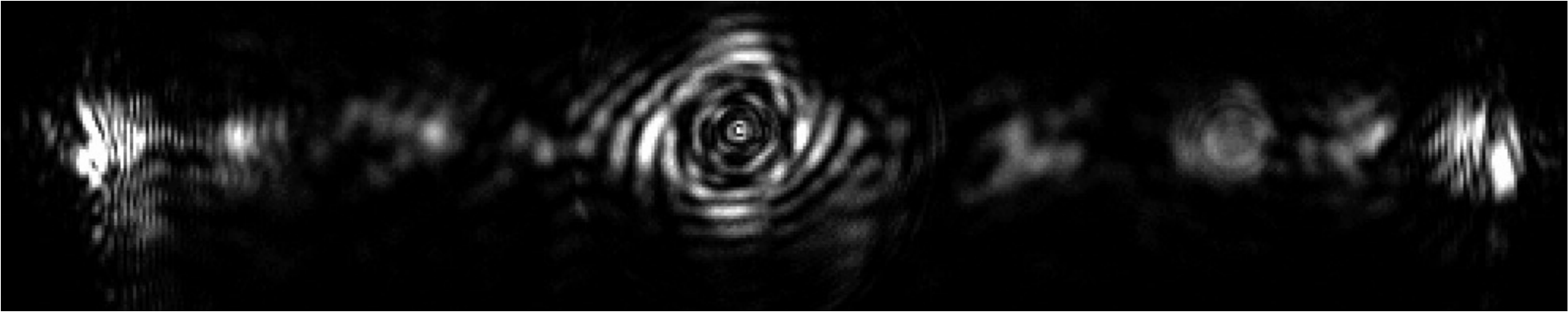}
        \caption{Best aligned focal-plane image of mask design without focus, designed to redistribute spots from a circular pupil on 4$\times$4 grid into a 1$\times$12 subaperture geometry.}
        \label{fig:nanoscribe-b}
    \end{subfigure}
    \begin{subfigure}[t]{0.195\textwidth}
        \includegraphics[width=1.0\textwidth]{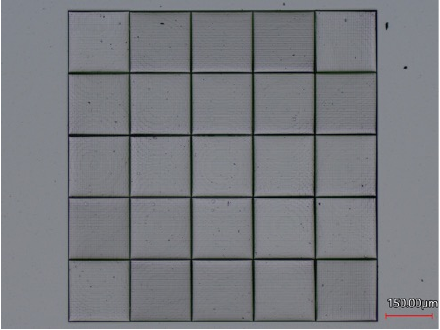}
        \caption{Microscope image of mask with focus (400~$\mu$m lenslet pitch).}
        \vspace{3pt}
        \label{fig:nanoscribe-c}
    \end{subfigure}
    \hspace*{2pt}
    \begin{subfigure}[t]{0.78\textwidth}
        \includegraphics[width=1.0\textwidth]{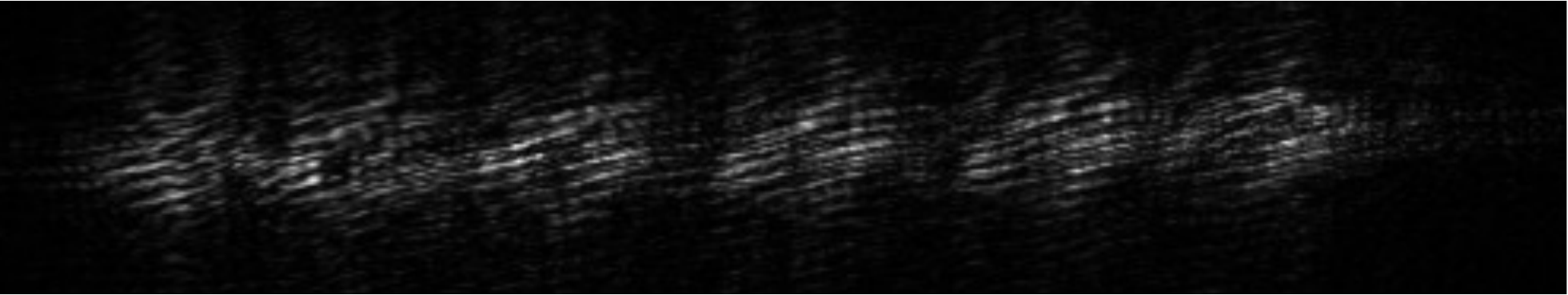}
        \caption{Best aligned focal-plane image of mask design with focus, designed to redistribute spots from a circular pupil on a 5$\times$5 grid into a 1$\times$21 subaperture geometry.}
        \label{fig:nanoscribe-d}
    \end{subfigure}
    \vspace*{3pt}
    \caption{Results from preliminary testing of fastrSHWFS masks fabricated at LLNL by Nanoscribe 3D printing and tested on a LLAMAS sub-bench.}
    \label{fig:nanoscribe}
\end{figure}
The results of Nanoscribe fastrSHWFS mask fabrication and testing is shown in Fig.~\ref{fig:nanoscribe}. Microscope images are shown instead of Zygo profilometry since we found that the high printing depths in these designs were not well-supported by the narrow range of fringes available to such interferometric characterization methods. The design with and without focus assumes a respective 8- and 6-bit depth discretization of a continuous design (depth discretization at or below 8 bits is required in the default mode for Nanoscribe printing, but see below), which has maximum respective printing depths of 12.7 and 5.2~$\mu$m, corresponding to respective minimum step sizes of 66 and 80~nm. Both designs use a spatial pixel size of 5~$\mu$m, corresponding to the techniques' diffraction-limit. This diffraction-limit nominally also limits depth step size to this resolution, but a piezo stage allows nanometer-level step sizes (also see below). 

The masks are aluminum coated and tested on a LLAMAS sub-bench at $\lambda$ = 635 nm, which  is setup with an adjustable diameter pupil stop conjugated to the fastrSHWFS mask (which is mounted on an XY micrometer-adjustable stage), which reflects the beam by $\sim20^\circ$ and is followed by an $F = 35$~mm, 1''$\varnothing$ imaging lens that, as we learned from Zeiss mask testing in \S\ref{sec:Zeiss}, is chosen and verified to not vignette the post-lenslet fastrSHWFS beam. The same CMOS camera on a Z micrometer stage as used in \S\ref{sec:Zeiss} is used in this setup to find the focus for the best quality spots. For the mask without focus, we found that a precisely aligned 400-$\mu$m pinhole mask (i.e., aligned with the XY micrometer adjusters mounting the fastrSHWFS mask at the $\sim10$ $\mu$m level) was needed to remove background signal on the flat, unpatterned substrate surface that we printed the mask on. The corresponding testing result shown in Fig.~\ref{fig:nanoscribe-b} shows our Nanoscribe masks with focus are not of sufficient quality to produce diffraction-limited spots, particularly at high angles where the mask printing depths are higher and the spot quality is clearly worse. The artifacts in the center of the image could be from imperfect collimation between our 400-$\mu$m-wide pupil stop and our mounted fastrSHWFS mask (i.e., below the sensitivity of the shear plate we used for alignment) and/or aforementioned non-zero reflection angle effects causing beam ellipticity ($\sim$6\% ellipticity for our $20^\circ$ reflection angle). The mask with focus in Fig.~\ref{fig:nanoscribe-c}-\ref{fig:nanoscribe-d} is also shown to not reach sufficient quality in a similar testing setup as described above, but with an adjustable iris pupil stop instead of a 400-$\mu$m pinhole. Similar effects as for the Nanoscribe mask without focus are likely cause degraded spot quality for the mask with focus, but with the added effect of discretization of the focus term in the fastrSHWFS mask. These results have indicated that a key area needed for improvement is to print Nanoscribe masks in continuous mode, requiring an STL file (a stereolithography CAD file format) to describe the continuous printing process. Such efforts are planned for further development in the near future.
\section{conclusions}
\label{Sec: conclusion}
We have presented the fastrSHWFS concept (\S\ref{sec:concept}) and motivated its benefit of reduced latency to extreme AO and future habitable exoplanet imaging systems (\S\ref{sec:motivation}). Our preliminary testing (\S\ref{sec:testing}) of both transmissive (\S\ref{sec:Zeiss}) and reflective (\S\ref{sec:nanoscribe}) versions of fastrSHWFS masks have not yielded diffraction-limited spot quality, although each section discusses potential areas for improvement as this concept will be further developed.
\acknowledgments 
This work was performed under the auspices of the U.S. Department of Energy by Lawrence Livermore National Laboratory under Contract DE-AC52-07NA27344. This document number is LLNL-PROC-865589.
%
\bibliography{report} 
\bibliographystyle{spiebib} 
\end{document}